# Broadband Multifocal Conic-Shaped Lens

Yanjun Bao[†], Qiao Jiang[†], Yimin Kang, Xing Zhu, and Zheyu Fang[*]

School of Physics, State Key Lab for Mesoscopic Physics, Peking University, Beijing 100871,

China

[†]These authors contributed equally to this work.

[*]Email: zhyfang@pku.edu.cn

Multifocal lens, which focus incident light at multiple foci, are widely used in imaging systems and optical communications. However, for the traditional design strategy, it combines several lenses that have different focal points into a planar integrated unit, resulting a low imaging quality due to the high background noise. Here, we propose two kinds of multifocal metalens with Au nanoslits arranged in an elliptical and a hyperbolic shape, which are able to effectively focus incident light at all of the foci with constructive interference, and extremely decrease the background noise and improve the lens imaging performance at the nanoscale. We further demonstrate that, the proposed metalens can possess a broadband operation wavelength changed from 600 nm to 900 nm, with its dual-polarity actively controlled by the incident circular polarization lights. With great agreement between the experimental and simulation results, our proposed conic-shaped metalens provides a significant potential for the future integrated nanophotonic device.

**INTRODUCTION**

Lens as one of fundamental optical components, has served people for hundreds years, shows marvelous applications in our daily life. With development of the modern industrialization, traditional lens fails in the miniaturization because of the optical diffraction limit and its long optical path for phase change accumulation. Metamaterials with designed patterns, on the other side, have been demonstrated effectively used to control the electromagnetic waves with custom-tailored optical response, such as negative index (*1, 2*) and cloaking (*3, 4*). Recently, metasurface was presented the ability to manipulate the wavefront of light, and introduce abrupt phase shift within a deep subwavelength distance at the interface (*5-13*). Ultra-thin metalens based on metasurface further breaks the thickness constraint of the traditional lens, and allows the miniaturization of future optoelectronic device (*14-20*).

Multifocal lens, which is used to focus incident light at multiple foci, has widely used in the imaging system and optical communications (*21-25*). In comparison with the conventional phase accumulation design (*21, 22*), metasurface engineering was recently implemented for the multifocal lens with the solution of designing several zone areas, and for each of the zone corresponds to one of the focal points (*23-25*). Though the light can be focused effectively for each of the zone, the entire multifocal lens shows a decreased optical performance, because the lens in a given region only constructively contributes to its corresponding focal point, and at the same time it also increases the background noise to the other areas. The strong background noise seriously impedes the application of multifocal lens for the ultrasensitive signal collection. Therefore, there is a huge demand for the design of efficient multifocal lens that can focus the incident light at all its focal points with constructive interference.

Metallic nanoslits arranged in an elliptical and a hyperbolic shape have been successfully demonstrated to have different and peculiar optical spin properties in the optical near-field (*26*). Due to different geometries, the ellipse- and hyperbola-shaped metasurfaces can realize optical spin Hall effect and spin-selective effect, respectively (*26*). In this letter, we further show that these two conic-shaped metasurfaces can not only realize the two optical spin phenomena in the far field, but also can be applied as the multifocal lens with constructively interfering at its all focal points. The polarity and focal positions of the metalens can be further controlled with the helicity of incident circular polarization (CP) light. The proposed metalens configuration provides a possible solution for the future multifocal lens design and easy fabrication.

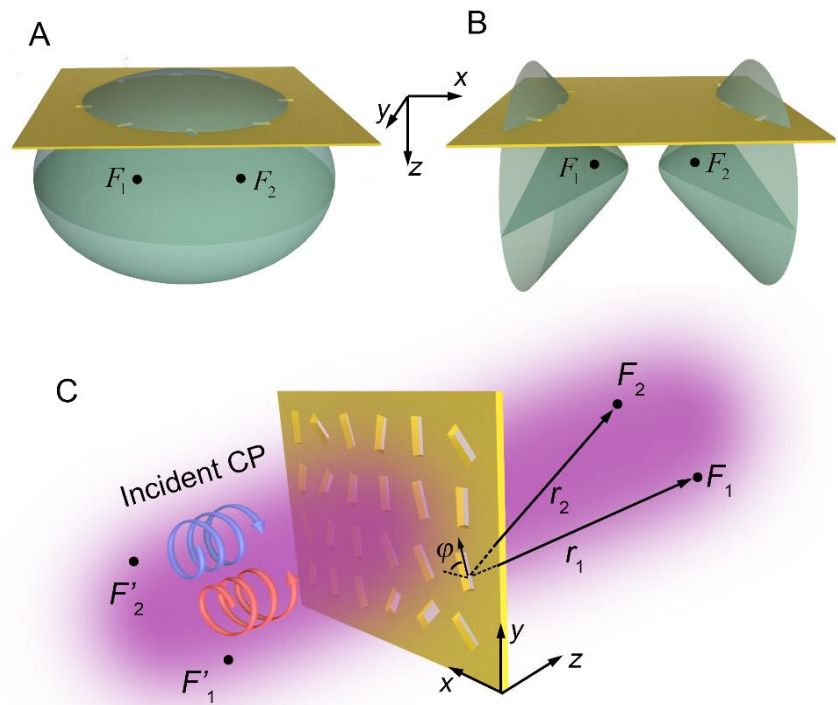

**Fig. 1. Conic-shaped metalens for constructively interference at multifoci.** (**A** and **B**) Schematics of an ellipsoid (A) and hyperboloid (B) with its two foci at points $F_1$ and $F_2$. The intersection of a plane with the ellipsoid (hyperboloid) results in an elliptical (hyperbolic) curve. (**C**) Schematic of a metalens composed of nanoslits with two foci at $F_1$ and $F_2$ under the illumination of CP light. $\varphi$ is the orientation angle of nanoslit

relative to the *x* axis. Points $F_1'$ and $F_2'$ are the mirror points of $F_1$ and $F_2$ respect to the metalens.

**RESULTS AND DISCUSSION**

We first consider an ellipsoid and a hyperboloid (Fig. 1A and 1B), which are defined by equation of $r_1 \pm r_2 = constant$, where $r_1$ and $r_2$ are the distances from the quadric surfaces to their two foci $F_1$ ($x_1$, $y_1$, $z_1$) and $F_2$ ($x_2$, $y_2$, $z_2$), respectively. To apply them into optics, we introduce a geometric-phase term $\sigma_\pm \psi$ that is distributed over the two quadric surfaces and rewrite their equations as follows:

$$(kr_1 + \sigma_\pm \psi) + (kr_2 + \sigma_\mp \psi) = constant \quad \text{(ellipsoid)} \quad (1)$$

$$(kr_1 + \sigma_\pm \psi) - (kr_2 + \sigma_\pm \psi) = constant \quad \text{(hyperboloid)} \quad (2)$$

where $k$ is the wave vector of light and $\sigma_\pm = \pm 1$ corresponds to the helicity (or spin) of right- (RCP) and left- (LCP) circular polarization. Assuming that the geometric-phase distribution satisfies that the phase term within the first bracket $kr_1 + \sigma_\pm \psi$ is a constant value (modulo $2\pi$) over the two quadric surfaces, the phase terms within the second bracket are automatically constant values, too. This means that light with different spin states can simultaneously constructively interfere at each of the two foci of the ellipsoid and light with one particular spin state can be accumulated (constructively interfering) at both foci of the hyperboloid. The two phenomena can be considered as optical spin Hall effect and spin selective effect. This difference is arisen from the opposite geometric-phase term in the second brackets in Eq. (1) and Eq. (2). The constructive interference at both foci is the property of an ideal multifocal lens. The intersection of a plane with the ellipsoid and hyperboloid results in an elliptical and a hyperbolic curve, as we called ellipse-shaped metalens (ESM, Fig. 1A) and hyperbola-shaped metalens (HSM, Fig. 1B).

Traditional method for constructing such a two-foci metalens is to design two area patterns with each of the area presenting one functionality (*23, 24, 27*), However, as mentioned before, such metalens cannot focus light constructively at both foci, therefore the background noise is high, and imaging quality is reduced. We expect that our proposed conic-shaped metalenses may decrease the background noise and improve the lens imaging performance, as we will demonstrate below.

To introduce the geometric phase, we note that with the illumination of LCP/RCP incidence, the transmission light through a nanoslit perforated at metallic film carries an additional geometric phase $\pm 2\varphi$ ($\psi=2\varphi$) in its cross polarization (*5, 6, 15, 28*), where $\varphi$ is the orientation angle of the nanoslit relative to the *x* axis, as shown in Fig. 1C. The orientation of each nanoslit of the ESM and HSM can be determined by the constraint equation of $kr_1+ 2\sigma_{\pm}\varphi$=constant (modulo $2\pi$).

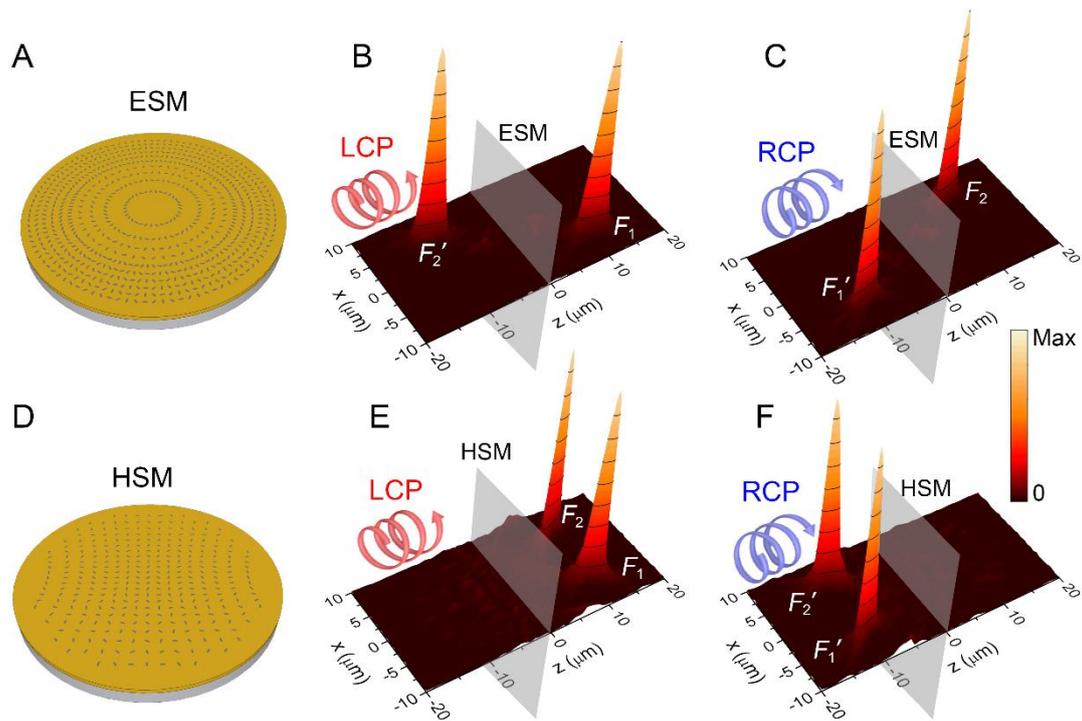

**Fig. 2. Numerical demonstrating of the focus of CP light in ESM and HSM.** (**A** to **C**) Schematics of the designed ESM (A) and simulated electric field intensities at *xz* plane (x∈[-10,10] μm, *y*=0 μm, z∈[-20,20] μm ) for LCP (B) and RCP (C) incidences, respectively. (**D** to **F**) The same as A-C, but for HSM. The CP light is incident from the left side and scattered to the right side of the metalens. The distance between the two focal points

($F_1$ and $F_2$) are 10 μm, and the focal length of the metalens is $f$=10 μm. The metalens is located at plane $z$=0 μm.

In order to verify our proposed ideas, ESM is designed that LCP and RCP light are focused at points $F_1$ and $F_2$, respectively ($2\varphi+kr_1=constant$ and $-2\varphi+kr_2=constant$), while HSM is designed that LCP light is focused at both foci $F_1$ and $F_2$ ($2\varphi+kr_1=constant$ and $2\varphi+kr_2=constant$). The two metalenses have a radius of 10 μm, and operate at wavelength of 785 nm (Fig. 2A, 2D and fig. S1). The coordinates of two focal points are $x_1=-x_2=-5$ μm, $y_1=y_2=0$ μm and $z_1=z_2=10$ μm. To demonstrate our prediction, finite-difference time-domain (FDTD) simulations were performed to calculate the electric field intensities at $xz$ plane with $y=0$ μm. Details of the calculation of the fields at any positions (including virtual plane) can be found in Methods. Figure 2 (B and C) shows the field intensity distributions of ESM with LCP (Fig. 2B) and RCP (Fig. 2C) incidences. As expected, LCP and RCP lights are focused at $F_1$ and $F_2$, respectively. Moreover, we also observed a virtual focal point $F_2'$ (mirror point of $F_2$) with LCP incidence and a virtual focal point $F_1'$ (mirror point of $F_1$) with RCP incidence. This is because the ESM also fulfills the following two equations: $-2\varphi-kr_1=constant$ and $2\varphi-kr_2=constant$, which correspond to the virtual focal points $F_1'$ and $F_2'$ with RCP and LCP incidences, respectively. The proposed ESM has dual polarity (positive and negative) with a real and a virtual focus under the illumination of CP light.

For the HSM, Fig. 2E shows that there are two focal points $F_1$ and $F_2$ at the real focal plane with the illumination of LCP light. When the polarization is altered to RCP (Fig. 2F), two virtual focal points $F_1'$ and $F_2'$ emerge at the virtual focal plane because the HSM also fulfills the two equations: $-2\varphi-kr_1=constant$ and $-2\varphi-kr_2=constant$. The polarity of HSM changes from positive with LCP incidence to negative with RCP incidence. Because the light is focused off-axis of the metalens, the intensity distribution at focal point is asymmetric (fig. S2). The simulated full width at half maximum (FWHM) of the focus along *x* axis for both metalens is about 560 nm, less than the incident wavelength 785 nm. The spot size and FWHM are strongly dependent on the radius of metalens and can be reduced by increasing the lens radius (fig. S3).

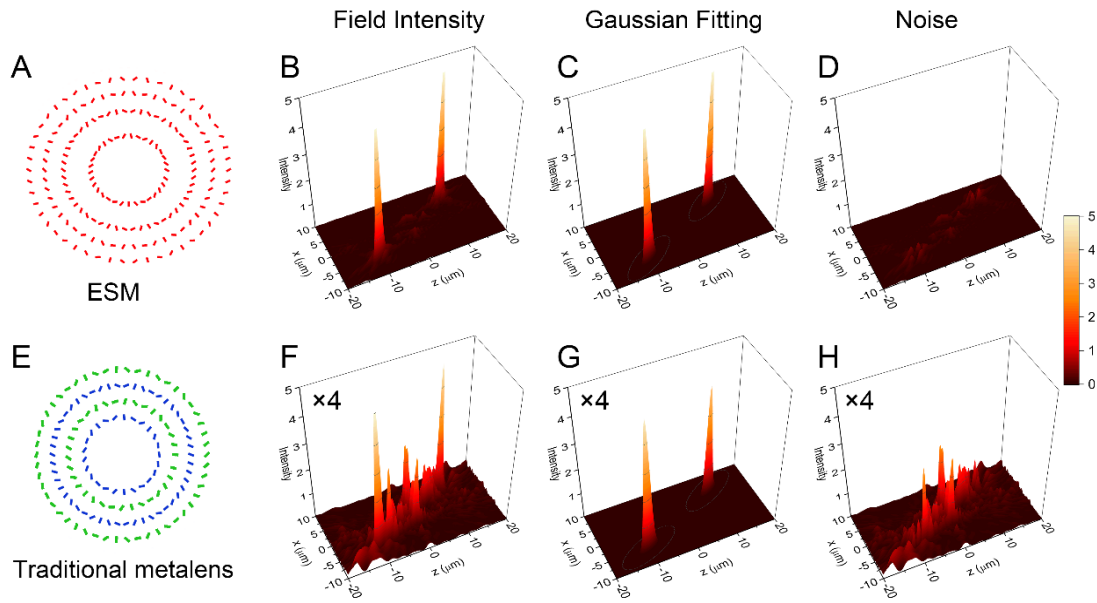

**Fig. 3. Comparison of focal performance between ESM and traditional metalens.** (**A**) Schematics of designed patterns of ESM. All the nanoslits in ESM constructively interfere at both foci. (**B** to **D**) Electric field intensity distribution (B), Gaussian fitting signals of the two focal peaks (C) and the noises (D) for ESM at *xz* plane with RCP incidence. (**E** to **H**) The same as A-D, but for traditional metalens. The traditional metalens is divided into two zones, with the green nanoslits contribute to one focus, and the blue nanoslits contribute to the other. The intensities in F, G and H are magnified by a factor of 4.

To compare with our conic-shaped metalens, we use traditional method to design a metalens (Fig. 3E), which includes two zones, with one (green nanoslits) corresponds to focus $F_1$ under LCP incidence and the other (blue nanoslits) corresponds to focus $F_2$ under RCP incidence. This traditional metalens has the same radius and the same number of nanoslits with ESM (Fig. 3A). The full designed patterns of ESM and traditional metalens are shown in fig. S1 and fig. S4, respectively. Figure 3 (B and F) shows the field intensity distribution at *xz* plane with RCP incidence for the two metalenses. Because only half-zone nanoslits constructively interfere at the foci, the peak intensity of traditional metalens is about one quarter of that of ESM. To estimate the signal-to-noise ratio (SNR), we use two-dimensional Gaussian function to fit the peaks as the signals (see the Supplementary Materials), as shown in Fig. 3 (C and G) for ESM and traditional metalens, respectively. The noise is defined as the absolute value of the difference between the total field intensity and the Gaussian fitting functions (Fig. 3, D and H). It is clearly observed that the background noise of traditional metalens is much larger than that of ESM. The calculated results show that the SNR of ESM is 2400, which is about four orders higher than that of traditional metalens (0.15).

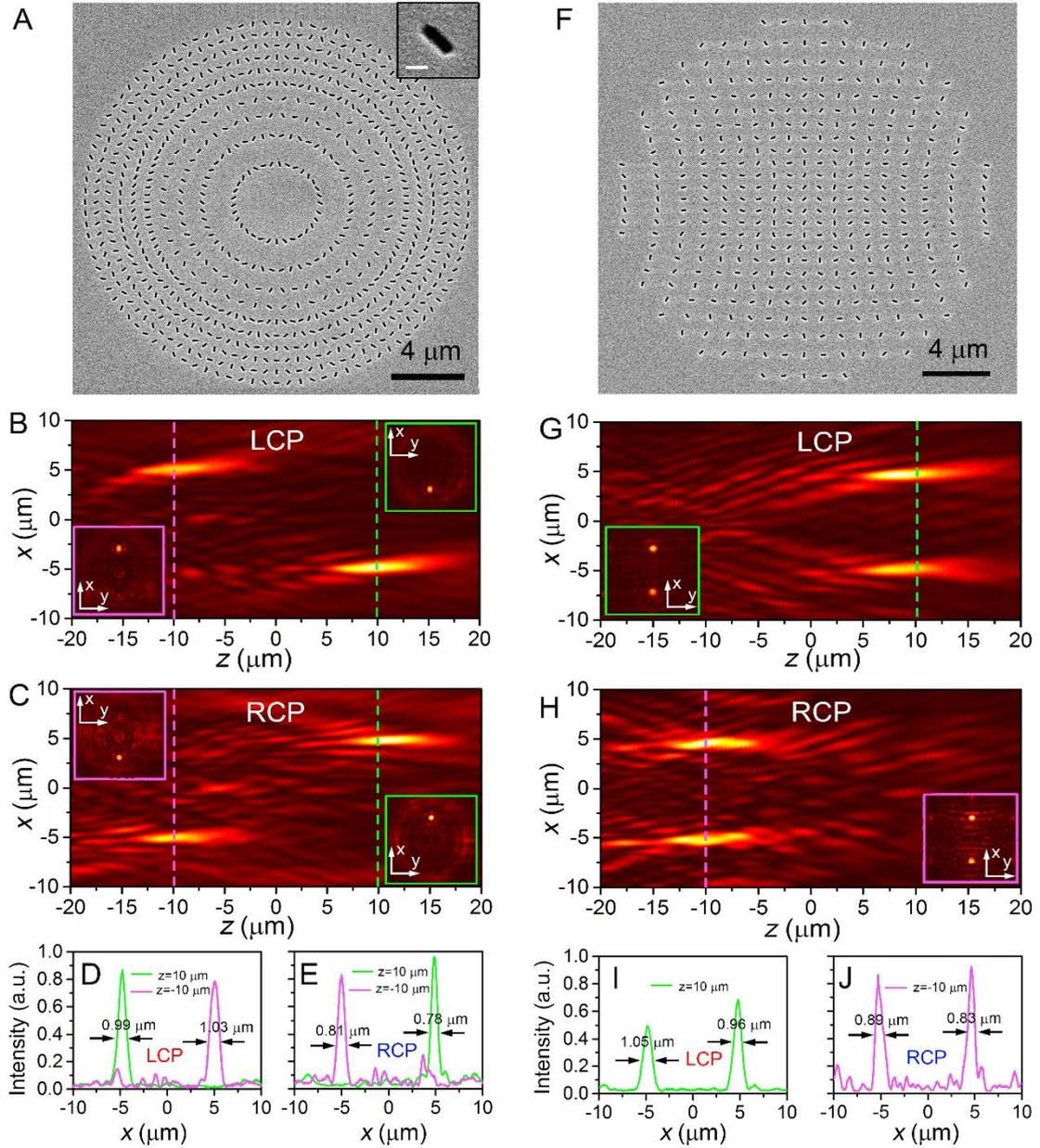

**Fig. 4. Experimental demonstration of the focusing performance for ESM and HSM**. (**A**) SEM images of ESM. Each nanoslit has a width of 100 nm and length of 300 nm. The inset shows a magnified SEM figure of a nanoslit (scale bar: 200 nm). (**B** and **C**) Measured $xz$ plane optical field intensities of ESM with LCP (B) and RCP (C) incidences. The insets shows the $xy$ planes at the real focal plane $z=10$ μm and virtual focal plane $z=-10$ μm. (**D** and **E**) Line plots of the intensities as a function of $x$ at real and virtual focal planes with LCP (D) and RCP (E) incidences for ESM. (**F** to **J**) The same as in (A-E) but for HSM.

To experimentally demonstrate the focusing performance of the designed metalens, we fabricated two conic-shaped metasurfaces by evaporating a 80 nm thickness of Au onto a glass substrate followed by focused ion beam (FIB) milling of the nanoslit pattern (See Methods). The parameters of the metalens are exactly the same as that in the FDTD simulations. Scanning electron microscopy (SEM) images of these two metalenses are shown in Fig. 4 (A and F). Far-field optical microscopy was employed to measure the optical intensity distributions (detection setup is shown in Methods and fig. S5). In the measurement, a motorize actuator was used to accurately adjust the distance between the objective lens and the sample. The optical intensity at *xz* plane can be obtained from a continuous images of *xy* planes at set distances over and below the sample.

Figure 4 (B and C) shows the measured optical intestines of ESM with LCP and RCP incidences at the wavelength of 785 nm. Two focal points with one real ($z=10$ μm) and one virtual ($z=-10$ μm) can be clearly observed for both cases, but with different focal positions. For the HSM, LCP light can be focused at two real focal points in front of the sample ($z=10$ μm), and RCP light is focused at two virtual focal points at the incident side ($z=-10$ μm), as shown in Fig. 4 (G and H). **Movie 1** and **Movie 2** show a gradual evolution of measured optical intensity at *xy* plane with CP light when *z* ranges from 20 μm to -20 μm for ESM and HSM, respectively. The measured FWHM of the focal points along *x* axis range from 0.78 to 1.03 μm for ESM (Fig. 4, D and E) and range from 0.83 to 1.05 μm for HSM (Fig. 4, I and J), which are comparable to the incident wavelength 785 nm, but larger than the simulation result 560 nm. This deviation may be induced by the nanofabrication and measurement inaccuracy.

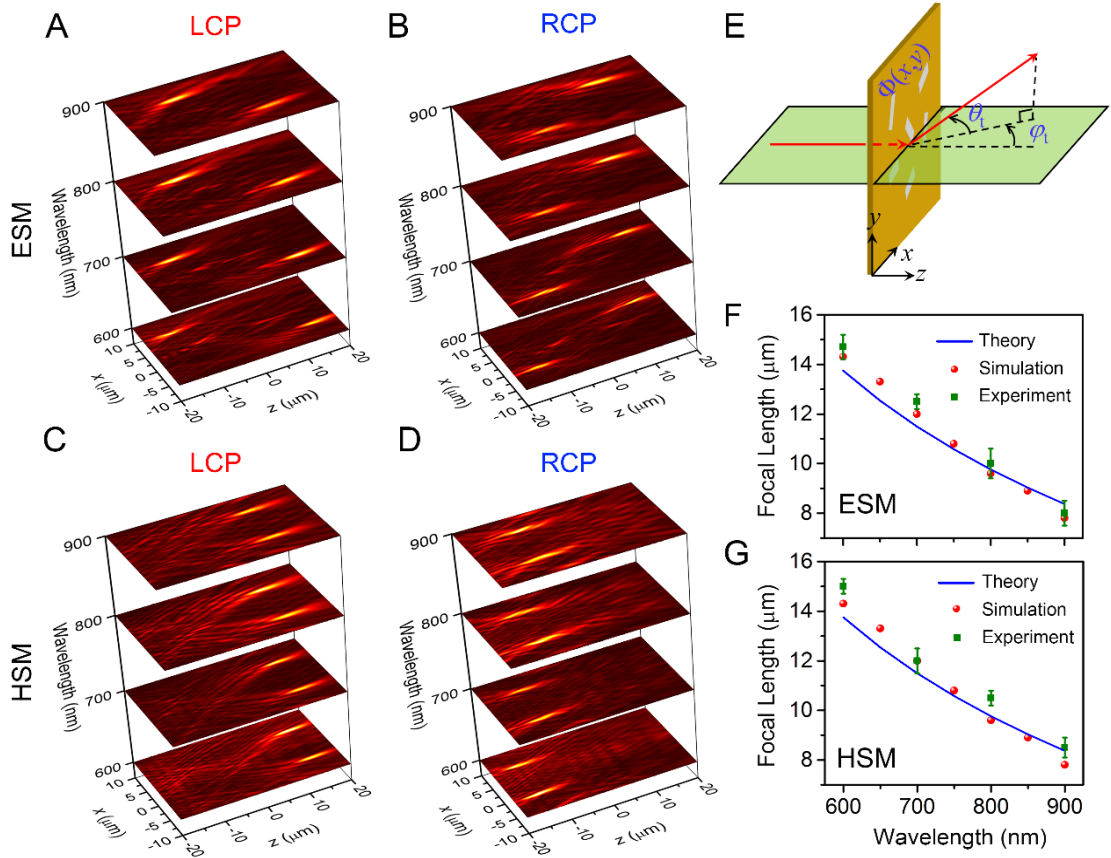

**Fig. 5. Broadband focal properties for ESM and HSM.** (**A** to **D**) Measured *xz* plane ($x \in$ [-10,10] μm, $z \in$ [-20,20] μm) optical intensities of ESM (A and B) and HSM (C and D) with LCP and RCP incident lights at wavelengths (*z* axis) of 600, 700, 800 and 900 nm. (**E**) Schematic of the defection of light incident on a metalens with a phase shift Φ(*x*,*y*). The light (red arrow) is normally incident and the transmission light is defected at angles with $\varphi_t$ and $\theta_t$. (**F** and **G**) Focal length as a function of wavelength for ESM (F) and HSM (G). The blue line, red solid circle and olive solid square show the theoretical predictions, simulated results and experimental results. For each metalens, there are four focal points in total with LCP and RCP incidences. The error bar is defined by the deviation of the focal lengths of the four foci.

We further demonstrate that our proposed conic-shaped metalenses can work well within a broadband wavelength ranging from 600 to 900 nm. Figure 5 (A to D) shows the measured optical intensities of *xz* planes for the two metalenses at selected wavelengths of 600, 700, 800 and 900 nm. Focal points can be clearly observed for all cases, which is consistent well with our simulation results in fig. S6. However, the position of the focus varies with the wavelengths for both metalenses. Figure 5 (F and G) are the simulated (red solid circle) and measured (olive solid square) results, which show that the focal length decreases with the increasing of the incident light wavelength. This phenomenon is associated with the intrinsic dispersion of optical lens, and can be explained by generalized Snell's laws of out-of-plane refraction (*7, 8*). Take the focal point $F_1$ for example, the phase shift $\varphi_s(x, y)$ induced by the metalens is

$$\Phi(x, y) = \frac{2\pi}{\lambda_d}(f - \sqrt{(x-x_1)^2 + y^2 + f^2}) \tag{3}$$

The phase gradient ($\frac{d\Phi}{dx}, \frac{d\Phi}{dy}$) at the metasurface deflects a portion of the normally incident light to an oblique transmission light that defined by two angles $\varphi_t$ and $\theta_t$, as shown in Fig. 5E. We therefore have (*7*)

$$\begin{cases} \cos\theta_t \sin\varphi_t = \frac{\lambda}{2\pi}\frac{d\Phi}{dx} \\ \sin\theta_t = \frac{\lambda}{2\pi}\frac{d\Phi}{dy} \end{cases} \tag{4}$$

Substituting Eq. 3 to Eq.4, the focal length can be theoretically calculated as cot($\varphi_t$)$x_1$ and shown as the blue lines in Fig. 5 (F and G), which also exhibits a decreasing trend of focal length with the increasing of the incident wavelength.

In summary, we have theoretically and experimentally demonstrated that the two conic-shaped metasurfaces can be used as multifocal lens in the far field. Their difference in the focusing performance is also discussed. In comparison with the traditional multifocal lens, our proposed metalens can focus the incident light at all of the focal points constructively and increase the SNR value by a factor of four orders of magnitude. We further show that these two conic-shaped metalenses can possess dual polarity under incident CP lights. Moreover, the designed metalens can operate well within a broadband incident wavelength from 600 to 900 nm in the visible range, and the dependence of focal length on incident wavelength was successfully explained by the generalized Snell's law. Because of the subwavelength size of nanoslit, the transmission efficiency of the metalens is relative low. One way to increase the transmission is to decrease the thickness of the metalens or to use dielectric block (*16*) instead to generate the geometric phase distribution. Our work successfully applies the conic concept into optics, and enables the miniaturization of multifocal lens with improved imaging performance.

**Methods:**

**Sample fabrication and measurement set-up.** We evaporated 80 nm thickness Au film onto a glass substrate and used focused ion beam (FIB) milling to etch the nanoslit structures on the film. The evaporation rate of Au was 0.5 Ås$^{-1}$ and the ion beam current was 20 pA to ensure the fabrication accuracy. Fig. S5 shows the schematic of the optical measurement setup. A super-continuum laser source (SC-5, YSLP) with wavelength ranging from 450 to 2000 nm is used as the incident light. An acousto-optic tunable filter (AOTF) is used to select a specific wavelength from the broadband laser source. The incident and transmitted CP lights were generated by a quarter-wave plate (QWP) and a polarizer on each side of the sample. The light scattered by the metasurface was collected with an objective (40×/0.75) and imaged on a CMOS camera. The sample was mounted on a three-dimensional stage with a motorize actuator (LTA-HS, Newport) to accurately adjust the distance between the objective and the sample with a resolution of 200 nm. All the optical components and the CMOS camera were positioned in a straight line.

**Numerical simulation.** We performed simulation by using a commercial software FDTD solutions (Lumerical). To calculate the electric field intensity at any positions, we just record the near field intensities at *xy* plane (large enough to collect most of the light) below the metalens ($z>0$). A near to far field projection was employed to calculate angular distribution of the far fields $E(k_x,k_y)$. The total field was obtained as the sum of all the plane waves multiplied by a phase factor: $E(x,y,z) = \sum E(k_x,k_y) e^{i(k_x x + k_y y + k_z z)}$. The virtual plane was defined as where $z<0$. This method enable us to calculate the field anywhere by just simulating a small region around the metalens.

# Acknowledgements

**Funding:** This work is supported by National Basic Research Program of China (973 Program, Grant No. 2015CB932403),National Science Foundation of China (Grant No. 61422501,11374023, and 61521004), Beijing Natural Science Foundation (Grant No. L140007),and Foundation for the Author of National Excellent Doctoral Dissertation of PR China (Grant No.201420),National Program for Support of Top-notch Young Professionals. **Author contributions:** Y.B. conceived the idea, performed the finite-difference time domain simulations, and fabricated the metalens. Y.B. and Q.J. characterized the samples and performed the measurements. Y.B., Y.K. and Z.F wrote the manuscript. All the authors discussed and analyzed the results. **Competing interests:** The authors declare that they have no competing interests. **Data and materials availability:** All data needed to evaluate the conclusions in the paper are present in the paper and/or the Supplementary Materials. Additional data related to this paper may be requested from the authors.

**REFERENCES AND NOTES**

1. Shalaev, V. M. Optical negative-index metamaterials. *Nat. Photonics* **1**, 41-48 (2007).
2. Smith, D. R.; Pendry, J. B.; Wiltshire, M. C. Metamaterials and negative refractive index. *Science* **305**, 788-792 (2004).
3. Liu, R.; Ji, C.; Mock, J. J.; Chin, J. Y.; Cui, T. J.; Smith, D. R. Broadband Ground-Plane Cloak. *Science* **323**, 366-369 (2009).
4. Schurig, D.; Mock, J. J.; Justice, B. J.; Cummer, S. A.; Pendry, J. B.; Starr, A. F.; Smith, D. R. Metamaterial electromagnetic cloak at microwave frequencies. *Science* **314**, 977-980 (2006).
5. Huang, L.; Chen, X.; Muhlenbernd, H.; Li, G.; Bai, B.; Tan, Q.; Jin, G.; Zentgraf, T.; Zhang, S. Dispersionless phase discontinuities for controlling light propagation. *Nano Lett.* **12**, 5750-5755 (2012).
6. Li, G.; Kang, M.; Chen, S.; Zhang, S.; Pun, E. Y.; Cheah, K. W.; Li, J. Spin-enabled plasmonic metasurfaces for manipulating orbital angular momentum of light. *Nano Lett.* **13**, 4148-4151 (2013).
7. Aieta, F.; Genevet, P.; Yu, N.; Kats, M. A.; Gaburro, Z.; Capasso, F. Out-of-plane reflection and refraction of light by anisotropic optical antenna metasurfaces with phase discontinuities. *Nano Lett.* **12**, 1702-1706 (2012).
8. Yu, N.; Genevet, P.; Kats, M. A.; Aieta, F.; Tetienne, J. P.; Capasso, F.; Gaburro, Z. Light propagation with phase discontinuities: generalized laws of reflection and refraction. *Science* **334**, 333-337 (2011).
9. Bao, Y. J.; Zu, S.; Zhang, Y. F.; Fang, Z. Y. Active Control of Graphene-Based Unidirectional Surface Plasmon Launcher. *ACS Photonics* **2**, 1135-1140 (2015).
10. Liu, L.; Zhang, X.; Kenney, M.; Su, X.; Xu, N.; Ouyang, C.; Shi, Y.; Han, J.; Zhang, W.; Zhang, S. Broadband metasurfaces with simultaneous control of phase and amplitude. *Adv. Mater.* **26**, 5031-5036 (2014).
11. Li, Z.; Palacios, E.; Butun, S.; Aydin, K. Visible-frequency metasurfaces for broadband anomalous reflection and high-efficiency spectrum splitting. *Nano Lett.* **15**, 1615-1621 (2015).
12. Yu, N.; Capasso, F. Flat optics with designer metasurfaces. *Nat. Mater.* **13**, 139-150 (2014).
13. Shitrit, N.; Yulevich, I.; Maguid, E.; Ozeri, D.; Veksler, D.; Kleiner, V.; Hasman, E. Spin-optical metamaterial route to spin-controlled photonics. *Science* **340**, 724-726 (2013).
14. Aieta, F.; Genevet, P.; Kats, M. A.; Yu, N.; Blanchard, R.; Gaburro, Z.; Capasso, F. Aberration-free ultrathin flat lenses and axicons at telecom wavelengths based on plasmonic metasurfaces. *Nano Lett.* **12**, 4932-4936 (2012).
15. Chen, X.; Huang, L.; Muhlenbernd, H.; Li, G.; Bai, B.; Tan, Q.; Jin, G.; Qiu, C. W.; Zhang, S.; Zentgraf, T. Dual-polarity plasmonic metalens for visible light. *Nat. Commun.* **3**, 1198 (2012).
16. Khorasaninejad, M.; Chen, W. T.; Devlin, R. C.; Oh, J.; Zhu, A. Y.; Capasso, F. Metalenses at visible wavelengths: Diffraction-limited focusing and subwavelength resolution imaging. *Science* **352**, 1190-1194 (2016).
17. Ding, X. M.; Monticone, F.; Zhang, K.; Zhang, L.; Gao, D. L.; Burokur, S. N.; de Lustrac, A.; Wu, Q.; Qiu, C. W.; Alu, A. Ultrathin Pancharatnam-Berry Metasurface with Maximal Cross-Polarization Efficiency. *Adv. Mater.* **27**, 1195-1200 (2015).
18. Khorasaninejad, M.; Chen, W. T.; Oh, J.; Capasso, F. Super-Dispersive Off-Axis Meta-Lenses for Compact High Resolution Spectroscopy. *Nano Lett.* **16**, 3732-3737 (2016).

19. Aieta, F.; Kats, M. A.; Genevet, P.; Capasso, F. Multiwavelength achromatic metasurfaces by dispersive phase compensation. *Science* **347**, 1342-1345 (2015).
20. Pors, A.; Nielsen, M. G.; Eriksen, R. L.; Bozhevolnyi, S. I. Broadband focusing flat mirrors based on plasmonic gradient metasurfaces. *Nano Lett.* **13**, 829-834 (2013).
21. Jia, J.; Zhou, C. H.; Liu, L. Superresolution technology for reduction of the far-field diffraction spot size in the laser free-space communication system. *Opt. Commun.* **228**, 271-278 (2003).
22. de Gracia, P.; Dorronsoro, C.; Marcos, S. Multiple zone multifocal phase designs. *Opt. Lett.* **38**, 3526-3529 (2013).
23. Chen, X.; Chen, M.; Mehmood, M. Q.; Wen, D.; Yue, F.; Qiu, C.-W.; Zhang, S. Longitudinal Multifoci Metalens for Circularly Polarized Light. *Adv. Opt. Mater.* **3**, 1201-1206 (2015).
24. Wang, W.; Guo, Z. Y.; Zhou, K. Y.; Sun, Y. X.; Shen, F.; Li, Y.; Qu, S. L.; Liu, S. T. Polarization-independent longitudinal multi-focusing metalens. *Opt. Express* **23**, 29855-29866 (2015).
25. Mehmood, M. Q.; Mei, S.; Hussain, S.; Huang, K.; Siew, S. Y.; Zhang, L.; Zhang, T.; Ling, X.; Liu, H.; Teng, J.; Danner, A.; Zhang, S.; Qiu, C. W. Visible-Frequency Metasurface for Structuring and Spatially Multiplexing Optical Vortices. *Adv. Mater.* **28**, 2533-2539 (2016).
26. Bao, Y.; Zu, S.; Liu, W.; Zhu, X.; Fang, Z. Revealing the spin optics in conic-shaped metasurfaces. *arXiv preprint arXiv:1602.01245* (2016).
27. Khorasaninejad, M.; Chen, W. T.; Zhu, A. Y.; Oh, J.; Devlin, R. C.; Rousso, D.; Capasso, F. Multispectral Chiral Imaging with a Metalens. *Nano Lett.* **16**, 4595-4600 (2016).
28. Huang, L.; Chen, X.; Bai, B.; Tan, Q.; Jin, G.; Zentgraf, T.; Zhang, S. Helicity dependent directional surface plasmon polariton excitation using a metasurface with interfacial phase discontinuity. *Light Sci. Appl.* **2**, e70 (2013).

# Supplementary Materials

**The PDF file includes:**

- Design of conic-shaped metalens

- Radius-dependent spot size

- Design of traditional metalens

- Calculation of signal-to-noise ratio (SNR)

- Schematic of the optical measurement setup

- Simulated results of the broadband focal properties of conic-shaped metalens

- fig. S1. Design patterns of conic-shaped metalens

- fig. S2. Simulated results of ESM and HSM at wavelength 785 nm

- fig. S3. Simulated spot size of focus as a function of the radius of metalens

- fig. S4. Metalens designed by traditional method

- fig. S5. Schematic of the optical measurement setup

- fig. S6. Simulated electric field intensity distributions of conic-shaped metalens within broadband wavelengths

## Design of conic-shaped metalens

The ESM is designed that LCP and RCP light are focused at points $F_1$ and $F_2$, respectively. We can write the two following equations: $2\varphi+kr_1=2\pi m+\varphi_1$ and $-2\varphi+kr_2=2\pi n+\varphi_2$, where $k$ is the wave vector of the incident light, $m$ and $n$ are two integers, $\varphi_1$ and $\varphi_2$ are constant values. By adding above two equations, we can have

$$r_1 + r_2 = (l+\varphi_0)\lambda \qquad \text{S1}$$

where $l$ is an integer, $\lambda$ is the wavelength of incident light, and $\varphi_0=(\varphi_1+\varphi_2)/2\pi$. The HSM is designed that LCP light is focused at both foci $F_1$ and $F_2$, which gives the two equations: $2\varphi+kr_1=2\pi n+\varphi_1$ and $2\varphi+kr_2=2\pi n+\varphi_2$. By subtracting these two equations, we have

$$r_1 - r_2 = (l+\varphi_0)\lambda \qquad \text{S2}$$

where $l$ is an integer, $\lambda$ is the wavelength of incident light, and $\varphi_0=(\varphi_1-\varphi_2)/2\pi$. Fig. S1 shows the pattern of our designed ellipse- and hyperbola-shaped metalens operating at wavelength of 785 nm. The focal length is 10 μm. For ellipse-shaped metalens (ESM), the $l$ value in Eq. S1 ranges from 28 to 37, and $\varphi_0=0$. Thus there are 10 elliptical curves in total. For hyperbola-shaped metalens (HSM), the $l$ value in Eq. S2 ranges from -9 to 9, and $\varphi_0=0$. Each nanoslit has a width of 100 nm, and length of 300 nm. The distance between two adjacent ones is chosen as 400 nm at least to avoid overlapping with each other. The radius of both metalenses are about 10 μm.

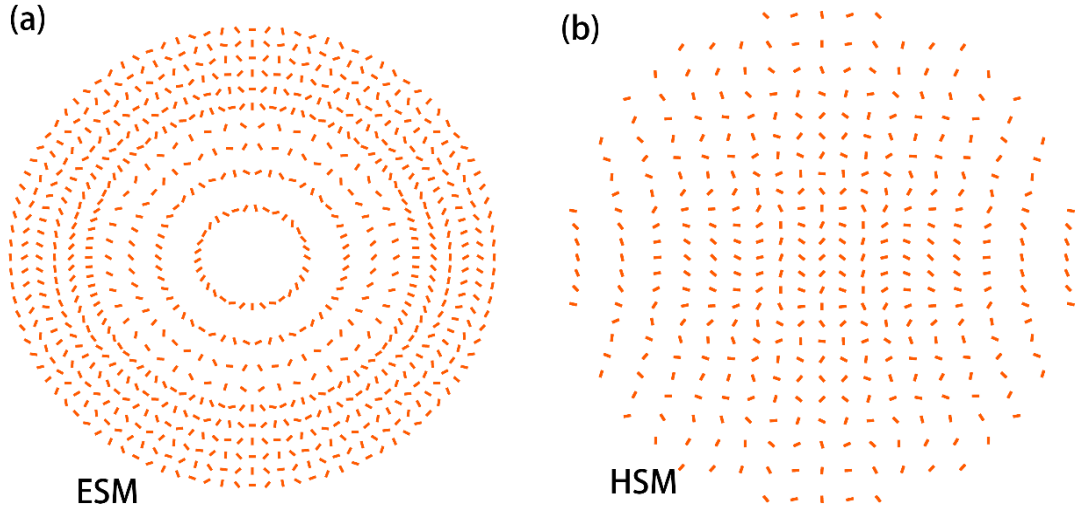

**fig. S1. Design patterns of conic-shaped metalens.** Top view of the patterns of ESM (a) and HSM (b).

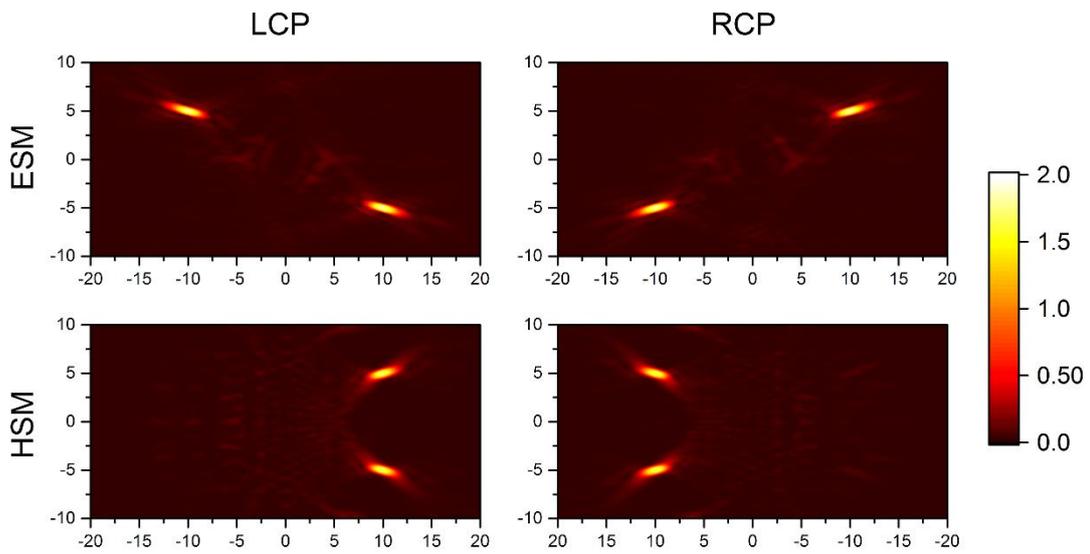

**fig. S2. Simulated results of ESM and HSM at wavelength 785 nm.** The data are the same as that of Fig. 2 in the main text but plotted in flat form. Because the light is focused off-axis, the resulted focus is asymmetric.

**Radius-dependent spot size**

The inset of Fig. S3 shows a focus of ESM. Because the focus is asymmetric, we define two FWHM values $w_1$ and $w_2$ along the mirror and major axes of focus. Fig. S3 shows the spot size of real focus of

ESM with RCP incidence as a function of radius of metalens. With the increasing of radius, the N.A. of metalens increases and the FWHM of spot size gradually decreases.

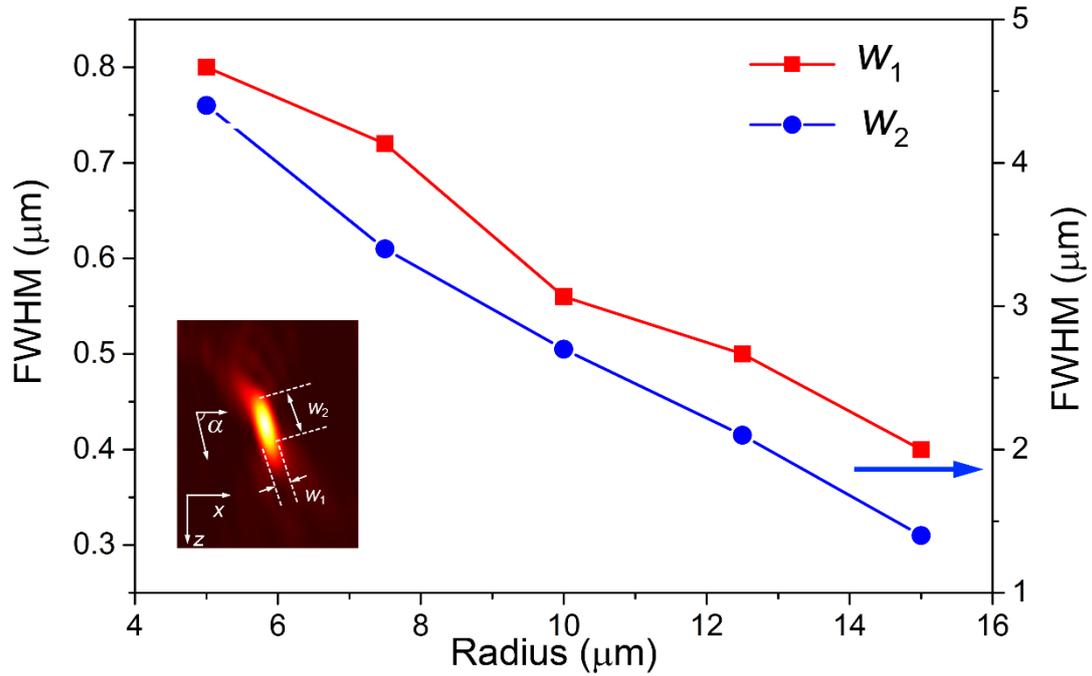

**fig. S3. Simulated spot size of focus as a function of the radius of metalens.** The inset shows the shape of the real focus, where the FWHMs $w_1$ and $w_2$ along the two different directions are defined. The operation wavelength is at 785 nm.

**Design of traditional metalens**

Fig. S4 shows the pattern of a metalens designed with traditional method. All the nanoslits are located on circular traces with different radii. The metalens are divided into two zones, with one area (green nanoslits) corresponding to $F_1$ ($x_1$=-5 μm, $y_1$=0 μm, $z_1$=10 μm) under LCP incidence and another (blue nanoslits) corresponding to $F_2$ ($x_2$=5 μm, $y_2$=0 μm, $z_2$=10 μm) under RCP incidence. The radius of the outer most boundary of metalens is 10 μm and the numbers of circles are 10, which is the same as ESM. With RCP incidence, there will be one real focus at $F_2$, and a virtual focus at $F_1$'.

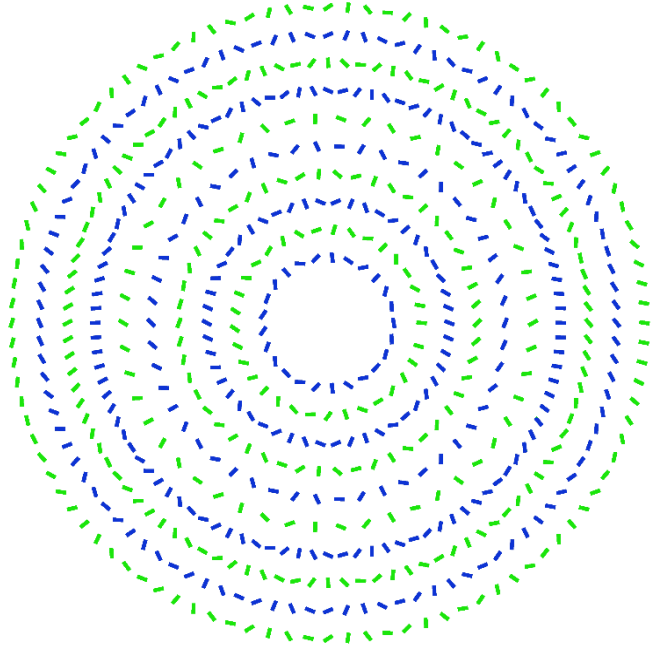

**fig. S4. Metalens designed by traditional method.** The metalens is divided into two zones (different colors), one corresponds to $F_1$ with LCP incidence and another corresponds to $F_2$ with RCP incidence.

## Calculation of signal-to-noise ratio (SNR)

With RCP illumination, both the ESM and traditional metalens have two focus spots, $F_2$ ($x_2,y_2,z_2$) and $F_1$' ($x_1$',$y_1$',$z_1$'). We first define a rotation matrix as

$$A(\alpha) = \begin{pmatrix} \cos(\alpha) & -\sin(\alpha) \\ \sin(\alpha) & \cos(\alpha) \end{pmatrix} \qquad \text{S3}$$

where $\alpha$ is the angle of the major axis of the spot relative to the $x$ axis (inset figure of Fig. S3).

To extract the signals, we fit the two peaks by two dimensional Gaussian function as

$$I = A e^{-\left[\frac{(x_n-x_{2n})^2}{(2w_1)^2} + \frac{(z_n-z_{2n})^2}{(2w_2)^2}\right]} + B e^{-\left[\frac{(x_n'-x_{1n}')^2}{(2w_3)^2} + \frac{(z_n'-z_{1n}')^2}{(2w_4)^2}\right]} \qquad \text{S4}$$

where the first and second terms are the fitting functions for foci $F_2$ and $F_1$', respectively, $A$ and $B$ are the magnitudes, $w_1$ ($w_3$) and $w_2$ ($w_4$) are the line widths along the mirror and major axes of foci, $\begin{pmatrix} x_n \\ z_n \end{pmatrix} = A(\alpha) \begin{pmatrix} x \\ z \end{pmatrix}$, $\begin{pmatrix} x_{2n} \\ z_{2n} \end{pmatrix} = A(\alpha) \begin{pmatrix} x_2 \\ z_2 \end{pmatrix}$, $\begin{pmatrix} x_n' \\ z_n' \end{pmatrix} = A(\beta) \begin{pmatrix} x \\ z \end{pmatrix}$, $\begin{pmatrix} x_{1n}' \\ z_{1n}' \end{pmatrix} = A(\beta) \begin{pmatrix} x_1' \\ z_1' \end{pmatrix}$, $\alpha$ and $\beta$ are the angles of the major axis of the foci $F_2$ and $F_1'$ relative to the $x$ axis.

The noise $N$ is defined as the absolute value of the difference between the total field $T$ and the Gaussian fitting signals $I$ as $N=\text{abs}(T-I)$. The total powers of signal and noise are the integration over the $xz$ plane $S$ (x∈[-10,10] μm, y=0 μm, z∈[-20,20] μm): $P_{signal} = \int IdS$ and $P_{noise} = \int NdS$.

The SNR is defined as

$$SNR = \frac{P_{signal}}{P_{noise}}.$$    S5

## Schematic of the optical measurement setup

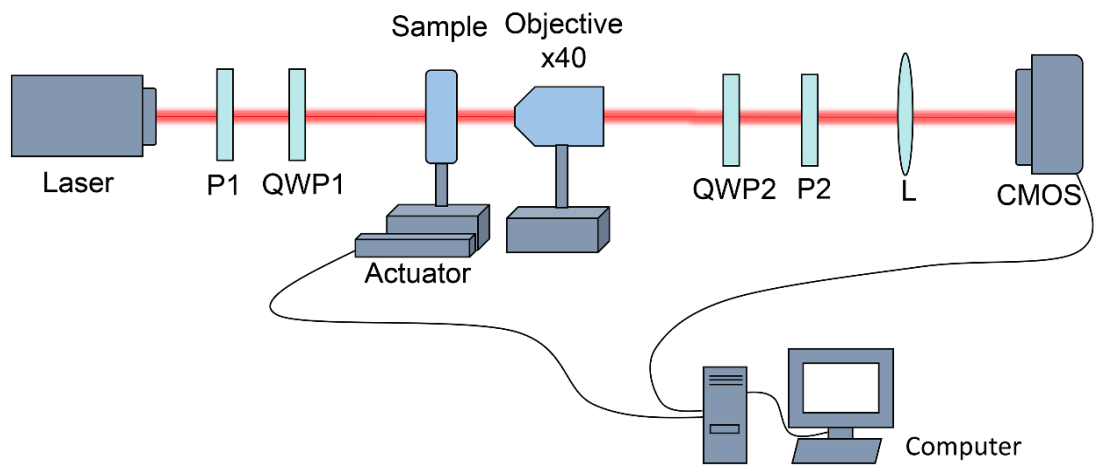

**fig. S5. Schematic of the optical measurement setup.** The incident and transmitted CP lights are generated by a quarter-wave plate (QWP) and a polarizer (P) on each side of the sample. The light scattered by the metasurface is collected with an objective (40×/0.75) and imaged on a CMOS camera. The sample is mounted on a three-dimensional stage with a motorize actuator (LTA-HS, Newport) to accurately adjust the distance between the objective and the sample with a resolution of 200 nm. All the optical components and the CMOS camera are positioned in a straight line.

**Simulated results of the broadband focal properties of conic-shaped metalens**

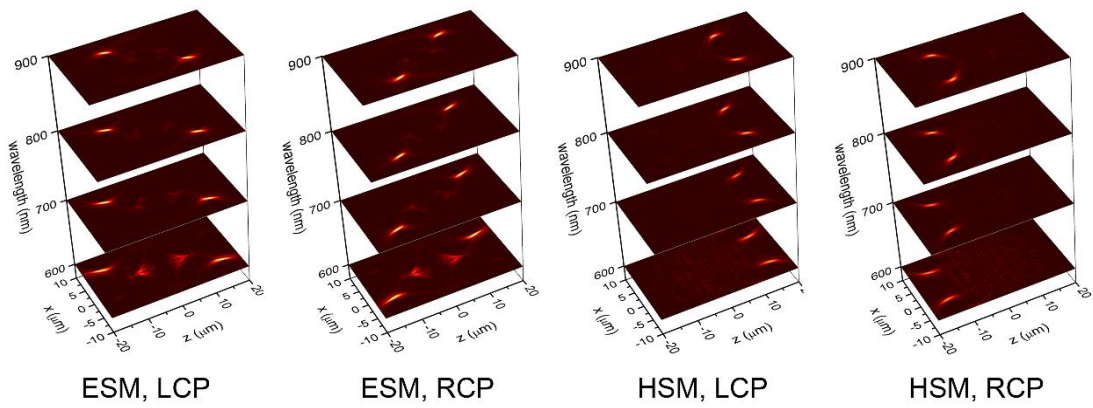

ESM, LCP      ESM, RCP      HSM, LCP      HSM, RCP

**fig. S6. Simulated electric field intensity distributions of conic-shaped metalens within broadband wavelengths.** The field intensities at *xz* plane (x∈[-10,10] μm, z∈[-20,20] μm) for ESM and HSM with LCP and RCP incident light at wavelengths of 600, 700, 800 and 900 nm. The simulation shows that the focal length gradually decrease with increasing of wavelength.